\documentstyle[aps,preprint,epsf]{revtex}
\begin{document}
\title{Nonequilibrium transport 
and population inversion 
in double quantum dot systems}
\author{Jun Zang$^{(1)}$, Joseph L Birman$^{(2)}$, 
A.R. Bishop$^{(1)}$, L. Wang$^{(1)}$ 
}
\address{
    $^{(1)}$Theoretical Division and Center for Nonlinear Studies, 
    Los Alamos National Laboratory, \\
    Los Alamos, NM 87545}
\address{
    $^{(2)}$Department of Physics, City College of CUNY, New York, N.Y. 10031}
\maketitle
\draft
\begin{abstract}
We present a microscopic theory for both equilibrium and 
nonequilibrium transport properties of coupled double quantum dots (DQD). A 
general formula for current tunneling 
through the DQD is derived
by the nonequilibrium Green's function method. 
Using a Hartree-Fock approach, 
effects of multi-level coupling and nonequilibrium electron
distributions in resonant tunneling 
are considered. We find that the peak in the resonant 
tunneling current through two symmetric dots
will split only  when the inter-dot coupling 
is stronger than dot-lead coupling. 
We predict that population inversion
can be achieved  in one dot in the 
nonequilibrium regime. 
\end{abstract}

Coupled quantum dot systems have received much attention recently
\cite{molenkamp,vaart95,waugh95,ruzin92,chen94,stafford94,niu-c,matveev95,golden95}.
Resonant tunneling through zero dimensional (0D) states of 
coupled quantum dots has been studied experimentally by 
several groups in both equilibrium ($|\mu_L-\mu_R|\ll k_BT$)
and nonequilibrium ($|\mu_L-\mu_R|\gg k_BT$) regimes  
\cite{molenkamp,vaart95,waugh95} (where $\mu_L$ and $\mu_R$ are
chemical potentials of leads externally attached to the 
quantum dots from left and right, respectively). The Coulomb blockade 
theory (CBT) was used to study the equilibrium properties 
of electronic tunneling in a double quantum dot (DQD) system 
\cite{ruzin92}, and more recently\cite{matveev95,golden95}, 
has been extended to explain the tunneling 
current peak splitting observed 
in the experiment of Waugh et al\cite{waugh95}.

Most previous theoretical studies on DQD systems
\cite{ruzin92,chen94,stafford94,matveev95,golden95} concentrated 
on the equilibrium properties. Interesting
properties in the nonequilibrium regimes \cite{vaart95}, where each dot 
can have {\it different} thermal nonequilibrium
states driven by the two leads, have received much less
attention.
In this paper, we use a microscopic
tunneling model to study the electronic transport through
DQD systems based on the nonequilibrium Green's 
function (NGF) method. Using
the NGF method, the resonant tunneling
current can then be derived exactly in both
equilibrium and nonequilibrium regimes for tunneling
through an interacting 
quantum dot \cite{meir92} and non-interacting multi-quantum
dots \cite{jz-dots}. Here we give a general expression for 
the current for tunneling 
through interacting multi-level double quantum dot systems.
The general current we derived can be used 
to obtain the well-known current equations in the equilibrium
regime. In the nonequilibrium regime, it is usually difficult to
satisfy detailed current balance between different dots and leads in
various approximations \cite{hershfield},
except in the case that the contribution to the imaginary 
self-energy is only from the lead-dot coupling. In this case,
the current equation is much simplified and can be written in
a compact form.

In this paper, we study resonant tunnelings through double
quantum dots, each of them having more than 50 electrons.
In these systems, only the quasi-particle states
are relevant to the resonant tunneling. The spectrum of quasi-particle
states near the Fermi surface will not change significantly
with changing the electron number from $N$ to $N+1$ except 
for a slight energy level shift mainly
due to the ``charging effect''. This has been confirmed for quantum dot
systems with $N>30$ by microscopic calculations \cite{lihong}.
In general, various scattering contributions to the imaginary
self-energy (scattering-in scattering-out processes) can be discussed
phenomenologically using our current equation below, which is 
useful for interpretation of experiments. We will use the 
general current equation derived here to discuss experimental
results performed in the nonequilibrium regime \cite{vaart95}; some of our explanations
are different from previous ones\cite{vaart95}.
In particular, we will see that the study of nonequilibrium 
electron distributions
in the quantum dots are crucial to understand nonequilibrium 
experiments \cite{vaart95}.
If the couplings of lead-dot and dot-dot
are $\Gamma$ and $|t|$, respectively, we predict
that in symmetric equilibrium experiments \cite{waugh95}
a peak of the resonant tunneling current
will split only when $|t| > \Gamma/2$
with a splitting magnitude $2(|t|^2-\Gamma^2/4)^{1/2}$.

Particularly, we predict that a population inversion
can be achieved in one of the two quantum dots when the tunneling between
the two dots is in resonance. This population inversion
should be easily realized under conditions
such as the low temperature
experiments of Ref.\cite{vaart95,waugh95}. At high temperatures,
the population inversion critically depends on electron-phonon (e-p)
or electron-electron (e-e) scattering induced relaxation times. 
This population inversion is similar 
to that in quantum cascade lasers (QCL) \cite{capasso94}
except that the former co-exists
with resonant tunneling and the tunneling is not
photon-assisted. Thus population inversion can be achieved
in the double quantum wells
under similar conditions as for QCL.

The structure of this paper is as follows:
in section~\ref{sec:model} we discuss the model and derive 
the general current equations; in section~\ref{sec:app} we 
discuss various approximations and the simplified current equations;
in section~\ref{sec:inversion} we discuss the non-equilibrium 
distribution of electrons in a double quantum dot system and
the possibility of population inversion;
in section~\ref{sec:exp}, we discuss two recent experiments;
and we conclude in section~\ref{sec:sum}.

\section{Model and Current Equations}
\label{sec:model}

We model the coupled DQD system with two attached leads 
by the Hamiltonian
\cite{lee-dot} 
$H = H_0 + H_t + H_I$,
with
\begin{eqnarray}
H_0 &=&  \psi_L^{\dagger}{\bf E}^L\psi_L+ 
\psi_R^{\dagger}{\bf E}^R \psi_R 
+ \sum_{i=1,2}\psi_i^{\dagger}{\bf E}^{(i)} \psi_i  
\nonumber \\
H_t &=& \psi_L^{\dagger} {\bf V_{L,1}} \psi_1 
+ \psi_R^{\dagger} {\bf V_{R,2}} \psi_2 
+ \psi_1^{\dagger} {\bf t_{12}} \psi_2 + h.c. ,
\label{eq:ham}
\end{eqnarray}
where the subscripts ($L$, $R$) represent the left- and right-lead, 
and ${i}$ sums over the two dots.
In Hartree-Fock (HF) representation,
$\psi^{\dagger}_{\alpha\in\{L,R\}} 
= (a^{\dagger}_{k_1\alpha},a^{\dagger}_{k_2\alpha},
\cdots,a^{\dagger}_{k_{N_{\alpha}}\alpha})$,
$\psi^{\dagger}_{i\in\{1,2\}} 
= (c^{\dagger}_{i,1},c^{\dagger}_{i,2},
\cdots,c^{\dagger}_{i,n_i})$.
Here \{$k_1,\cdots k_{N_{\alpha}}$\} and $\{1, 2,\cdots n_i\}$ are labels of states
in the lead and dot, $N_\alpha$ and $n_i$ are the
numbers of total states in the lead-$\alpha$ and dot-$i$, respectively.
The ${\bf E}^{i}$'s are diagonal matrices whose elements are 
calculated in the HFA. 
$H_t$ is a tunneling Hamiltonian.
${\bf V}_{\alpha,i}\equiv [V^{\alpha,i}_{k_m,l}]$ 
is the  ($N_{\alpha} \times n_{i}$)
hopping integral matrix; the matrix element $V^{\alpha,i}_{k_m,l}$
is the hopping integral from state $k_m$ in lead-$\alpha$ to
state-$l$ in dot-$i$. ${\bf t}_{12}\equiv [t^{1,2}_{l,m}]$ is a ($n_1 \times n_2$) matrix, and
the matrix element $t^{1,2}_{l,m}$
is the hopping integral from state-$l$ in dot-$1$ to
state-$m$ in dot-$2$.
$H_I$ describes residual interactions
such as electron-phonon (e-p) scattering, disorder
scattering, and e-e correlations.

Since the system under investigation is in non-thermal equilibrium
in the nonlinear region (with finite bias voltage), The natural
technique to use is that of non-equilibrium Green functions 
\cite{ngf,keldysh,su-ctgf} (NGF).
The NGF in the DQD in matrix form is defined by:
\begin{equation}
{\bf G}_{i,j}(\tau,\tau') = 
-i \left\langle {\cal T}_C\left\{ \psi_i(\tau)\psi^{\dagger}_j(\tau')
\right\} \right\rangle ,
\end{equation}
where ${\cal T}_C$ is the chronological time-ordering
operator on the contour $C$ of the closed time path, and
$\tau \equiv (\beta,t)$ with $\beta =\pm$ in the plus (minus) branch
of the contour $C$. 
We choose the bases as HF states, 
so the evolution of $\psi_i$ is governed by $H_t$ and $H_I$. 
Similarly, we define the GF for the two leads ${\bf G}_L$ and
${\bf G}_R$, and the ``off-diagonal'' NGF ${\bf G}_{L,1}$ and
${\bf G}_{R,2}$, etc. 
The four combinations of the time branches $(\pm,\pm)$ give
four linearly dependent component NGFs. 
Here we have used the notation that 
${\bf G}_{\gamma,\delta}$ without superscript index means 
a ($2\times 2$) matrix of ${\bf G}_{\gamma,\delta}^{\pm\pm}$
with superscripts. After a rotation in the ($2\times 2$) space,
we can obtain three independent
components of the NGF, i.e., 
${\bf G}^{r,a,+-}_{ij}$ (retarded, advanced and 
statistical component carrying information 
of nonequilibrium distributions). Another frequently used
(Keldysh) statistical component of NGF instead of
${\bf G}^{+-}$ is 
${\bf G}^{K}=2{\bf G}^{+-}+{\bf G}^{r}-{\bf G}^{a}$.
Each of the three components
is itself a ($n_i \times n_j$) matrix, where $n_i$ is the number
of total states in dot-$i$ or lead-$i$. (Note that we have 
used the notation $n_i$ for the leads instead of $N_{\alpha}$, this is only
for the convenience of description here).
The matrix elements of these $n_i \times n_j$ NGF matrices
 are just the scalar NGF,
e.g. 
$[G_{1,2}^{r}]_{l,m}(t,t')=
-i\Theta(t,t')\langle 
\{c^{\phantom{\dagger}}_{l,1}(t),c^{\dagger}_{m,2}(t')\}\rangle$.

Since the leads are open systems, we can treat them
as two equilibrium Fermi seas of non-interacting quasi-particles
with different chemical potentials. The degrees-of-freedom
of the leads can be integrated out and 
the self-energies (in matrix form) of the two
quantum dots due to the tunneling between the leads and dots
become \cite{jz-dots}:
${\bf \Sigma}^0_1(\omega) = 
{\sigma}_x{\bf V_L}^{\dagger}{\bf G}_L{\bf V_L}{\sigma}_x$;
${\bf \Sigma}^0_2(\omega) = 
{\sigma}_x{\bf V_R}^{\dagger}{\bf G}_R{\bf V_R}{\sigma}_x$,
where $\sigma_x$ is the Pauli matrix operating only in 
the ($2\times 2$) Keldysh space \cite{keldysh}.
Defining $i{\bf \Gamma}_{j}=
({\bf \Sigma}^0_{j})^r-({\bf \Sigma}^0_{j})^a$,
it is easy to see that:
$({\bf \Sigma}^0_1)^{+-}(\omega) = if_L(\omega){\bf \Gamma}_{1}(\omega)$
and
$({\bf \Sigma}^0_2)^{+-}(\omega) = if_R(\omega){\bf \Gamma}_{2}(\omega)$,
with $f_L(\omega)$ and $f_R(\omega)$ the Fermi-Dirac
distribution functions of the left and right leads, respectively.
Physically, the diagonal matrix element $[\Gamma_i]_{l,l}$
of ${\bf \Gamma}_{i}$ is the level broadening of state-$l$
in dot-$i$ due to the tunneling to all the states
in the neighboring lead.

The current equations can be derived formally using the 
Dyson equations. In our system, 
the Dyson equations (in Keldysh space) \cite{keldysh} 
can be written as:
\begin{eqnarray}
{\bf g}_i^{-1} {\bf G}_{ij} &=& {\bf \delta}_{ij}+{\bf \Sigma}_{ij}{\bf G}_{ij}
+{\sigma}_x{\bf t}_{12}{\bf G}_{\bar{\imath} j},
\label{eq:dyson}
\end{eqnarray}
where  ${\bf g}_{i,j}\equiv {\bf g}_i \delta_{i,j}$
is the {\it bare} GF for $H_0$, $\bar {\imath}\in \{\bar{1}\equiv 2, \bar{2}\equiv 1\}$.
To take into account the self-energy parts from 
the e-e correlations and e-p scattering $H_I$, $\bbox{\sigma}_{ij}$,
we obtain the self-energies 
${\bf \Sigma}_{ij} = {\bf \Sigma}_i^0 + \bbox{\sigma}_{ij}$.
If there is no coupling between $\psi_1$ and $\psi_2$
other than tunneling, then $\bbox{\sigma}_{12}=\bbox{\sigma}_{11}$ and
$\bbox{\sigma}_{21}=\bbox{\sigma}_{22}$. 

The Dyson
equations (\ref{eq:dyson}) can be solved easily:
${\bf G}_{ii}^{r, a} =
[ ({\bf g}_i^{r, a})^{-1} -
{\bf \tilde{\Sigma}}^{r, a}_i ]^{-1}$, 
${\bf G}_{ii}^{+-} = {\bf G}_{ii}^r 
{\bf \tilde{\Sigma}}^{+-}_i {\bf G}_{ii}^a$,
where we have used the total self-energy 
${\bf \tilde{\Sigma}}_i={\bf \Sigma}_{ii}+{\bf \delta\Sigma}_i$ with
\begin{eqnarray}
{\bf \delta\Sigma}_i &= & {\sigma}_x
{\bf t}_{12} \left[ {\bf g}_i^{-1} - 
{\bf \Sigma}_{i\bar{\imath}} \right]^{-1} {\bf t}_{12}^{\dagger}{\sigma}_x
\equiv {\sigma}_x{\bf t}_{12} {\bf \tilde{g}}_i {\bf t}_{12}^{\dagger}{\sigma}_x.
\label{eq:gt}
\end{eqnarray}
Here we have used five different self-energies. Their physical
meaning is following:
as described above, ${\bf \Sigma}^0_i$ is the self-energy due to
tunneling to the neighboring lead; $\bbox{\sigma}_{ij}$ is the
self-energy due to interactions $H_I$ without tunneling between
quantum dots; ${\bf \delta\Sigma}_i$ is the self-energy due to
tunneling to another dot (Note this self-energy is calculated
using the renormalized  GF ${\bf \tilde{g}}_i$); finally, 
${\bf \tilde{\Sigma}}_i$ is the total self-energy.

From the Dyson equations, the current can easily be calculated
at different dots:
\begin{mathletters}
\label{eq:curr}
\begin{eqnarray}
J^{(\alpha,j)}&=&(-1)^{j+1}{ie\over h}\int d\omega Tr\{ 
\left[({\bf \Sigma}^0_j)^r-({\bf \Sigma}^0_j)^a \right]
{\bf G}_{jj}^{+-}
\nonumber \\
&-&({\bf \Sigma}^0_j)^{+-} 
\left[{\bf G}_{jj}^r-{\bf G}_{jj}^a \right]
\}
\label{equationa}
\\
J^{(j)} &=& (-1)^j {ie\over h}\int d\omega Tr\{ 
\left[({\bf \delta\Sigma}_j)^r-({\bf \delta\Sigma}_j)^a \right]
{\bf G}_{jj}^{+-}
\nonumber \\
&-&({\bf \delta\Sigma}_j)^{+-}
\left[{\bf G}_{jj}^r-{\bf G}_{jj}^a \right]
\},
\label{equationb}
\end{eqnarray}
\end{mathletters}
where the superscripts $(\alpha,j)\in \{(L,1), (R,2)\}$ 
denotes the current tunneling 
from the lead to the dot-$i$, and $(j)$ is the inter-dot current  
calculated at dot-$j$. 

\section{Simplified Current Equations: Approximations}
\label{sec:app}

In general,
the introduction of 
$\bbox{\sigma}_{ij}$ makes the current in Eq.(\ref{eq:curr})  
non-conserved \cite{hershfield}. To obtain a current obeying detailed balance, 
current-conserving approximations for $\bbox{\sigma}$ are required.
However, since we adopt HFA here, when $H_I$ is neglected 
then the current in Eq.(\ref{eq:curr}) is automatically conserved
\cite{jz-dots}. The current is now written in a simpler form:
\begin{eqnarray}
J&=&{e\over h}\int d\omega (f^{FD}_L-f^{FD}_R) T_r\left[
{\bf G}_{22}^a 
 {\bf \Gamma}_1 
{\bf G}_{22}^{r}
{\bf t}_{12} {\bf \tilde{g}}^r_1 
{\bf \Gamma}_2
{\bf \tilde{g}}^a_1{\bf t}_{12}^{\dagger}
\right] ,
\label{eq:curr2}
\end{eqnarray}
where $f^{FD}_{L,R}$ is the Fermi-Dirac
distribution function and 
${\bf \tilde{g}}_i$ is defined in Eq.(\ref{eq:gt}).
Recall that 
${\bf t}_{12} {\bf \tilde{g}}^r_1 
{\bf \Gamma}_2
{\bf \tilde{g}}^a_1{\bf t}_{12}^{\dagger}$
is just the imaginary part of self-energy 
${\bf \delta\Sigma}_i$ due to tunneling between quantum dots.
The current equation (\ref{eq:curr2}) is very similar
to that for a single dot system, i.e. replace ${\bf \Gamma}_{\alpha}$ due to
tunneling to the lead by ${\bf \Gamma}_{ij}
=(\delta{\bf \Sigma}_{j})^r-(\delta{\bf \Sigma}_{j})^a$ due to tunneling 
from dot-$i$ to another dot-$j$.

The electron-electron (e-e) and electron-phonon (e-p) 
interactions have two kinds of effects:
renormalizing ${\bf E}^{(i)}$ due to $\Re \bbox{\sigma}^{r,a}$
and changing the distribution of electrons due to $\bbox{\sigma}^K$
(and $\Im \bbox{\sigma}^{r,a}$). In the former effect, we simply renormalize \cite{note3} 
${\bf E}^{(i)}$, so the current equation is the same as Eq.(\ref{eq:curr2}).
After we include the higher order (than HFA) energy shift, 
Eq.(\ref{eq:curr2}) contains rich correlation effects. 
One example
is for the two-impurity model in the Hartree approximation studied
by Niu et al \cite{niu-c}. One might attempt to calculate the interacting
NGFs using higher order (than Hartree approximation) 
truncation of the EOM and then calculate tunneling currents using
Eq.(\ref{eq:curr2}). This is questionable for double dot systems
except when the truncated EOM is in the form of a Dyson equation. 
Otherwise,
one needs to re-derive the current equation using the truncated
EOM.
For large quantum dots with more than $40$ electrons, 
only the quasiparticle
states near the Fermi surface are relevant to the resonant tunneling.
The spectral functions of these quasiparticle states do not change
significantly with adding/subtracting a couple of electrons in the 
dots except for 
a slight energy level shift due to the ``charging effects''. 
This has been
confirmed by microscopic calculations by Wang, Zhang, and Bishop 
\cite{lihong}, who also found that the many-electron energy levels 
are approximately equally spaced for quantum dots with
more than 30 electrons. Thus for quantum dots with a large occupation
numbers,
the energy shift due to e-e interactions can be treated 
phenomenologically (see Sec.\ref{sec:exp})
and the level spacing can be extracted from experiments.
The scattering-in/out effect due to $\bbox{\sigma}^K$
is difficult to treat correctly in various approaches
due to the detailed current balance problem mentioned above. 
However, qualitative effects can still be obtained after simplifying
Eq.(\ref{eq:curr}) appropriate to different conditions, 
in a similar fashion to
our discussions of nonequilibrium distributions below.

Up to now, our equations are general for multi-level dots. 
To study a specific system, we need to know the matrix
elements $[{\bf t}_{12}]_{l_i l_j}$ (hopping) and 
$\left[{\bf \Gamma}_i\right]_{l_i l_j}$. 
Let us first examine
the relevance of various matrix elements. The elements
$\left[{\bf \Gamma}_i\right]_{l_i l_j}$ for $l_i\neq l_j$ 
are due to tunneling from different levels $l_i$ and $l_j$
to the same state in the lead. These are usually not small
compared to the diagonal ones, and they describe the mixing of
levels in the dots due to tunneling to the leads. 
If we assume the energy of the states in the tunneling
window (i.e. $\mu_1 <E_i <\mu_2$) is much smaller than the
barrier height between leads and dots, then
$\left[{\bf \Gamma}_i\right]_{l_i l_j} = \Gamma_i$.
The matrix elements $[{\bf t}_{12}]_{l_i l_j}$ describe
wavefunction overlaps of different levels between two quantum
dots. The value of these matrix elements are different but
roughly of the same magnitude if their energies are comparable.
Similarly, if we assume the energy of the states in the tunneling
window (i.e. $\mu_1 <E_i <\mu_2$) is much smaller than the
barrier height between two dots, then
we can use the following assumption
$[{\bf t}_{12}]_{l_i l_j} = t$ 
(thick-shell model \cite{golden95}).
Using the assumption $\left[{\bf \Gamma}_i\right]_{l_i l_j} = \Gamma_i$, 
\begin{eqnarray}
\left[ {\bf \tilde{g}}^{r,a}_{\alpha}\right]_{ij}= \cases{
{1\over w_i^{r,a}}\left(1+{1 \over w_i^{r,a}} 
{(\Sigma^0_{\alpha})^{r,a} \over 1-
(\Sigma^0_{\alpha})^{r,a} \sum_k {1\over w_k^{r,a}} } \right)
& $i=j$; \cr
\cr
{(\Sigma^0_{\alpha})^{r,a} \over
w_i^{r,a} w_j^{r,a} \left(1-(\Sigma^0_{\alpha})^{r,a} 
\sum_k {1\over w_k^{r,a}}  \right) }
& $i\neq j$. \cr
}
\label{eq:case}
\end{eqnarray}
where $w_i^{r,a}=[({\bf g}_{\alpha}^{r,a})^{-1}]_{ii}$.

In the experiments \cite{vaart95,waugh95}, 
$\Gamma_i$, $|t| \ll \Delta E$ (spacing between energy
levels in the dots).
Using Eq.(\ref{eq:case}), it is easy to show 
that near the resonance $\omega \sim \epsilon_{l_0}$: 
for $l_i =l_0$, $\left[ {\bf \tilde{g}}^r_1\right]_{l_i l_i} 
\sim 1/\Gamma_1$;
for $l_i\neq l_0$; $\left[ {\bf \tilde{g}}^r_1\right]_{l_i l_i}
 \sim 1/[(l_i-l_0)\Delta E]$;
for $l_i\neq l_j\neq l_0$, $\left[ {\bf \tilde{g}}^r_1\right]_{l_il_j}
\sim (\omega-\epsilon_{l_0})/[(l_i-l_0)(l_j-l_0)(\Delta E)^2]$; for $l_i=l_0$, $l_j\neq l_0$,
$\left[ {\bf \tilde{g}}^r_1\right]_{l_il_j}
\sim 1/[(l_j-l_0)\Delta E]$. The properties of the
$[{\bf G}_{ij}^r]_{l_i l_j}$ are similar. 
Thus, at $\Gamma_i, |t| \ll \Delta E$, only the tunnelings between
levels close to each other are important.

For the resonant peak structure, we can simplify 
Eq.(\ref{eq:curr2}) near resonance,
$\mu_L >\epsilon^{(1)}_{l_1} \sim \epsilon^{(2)}_{l_2}
>\mu_R$,
by taking into account only the resonance level contribution:
\begin{eqnarray}
J & \simeq & \int d\omega 
 (f_L-f_R) {\Gamma_2 \Gamma_{12} \over
(\omega-{\epsilon}^{(2)}_{l_2}-\delta \epsilon^{(2)})^2+
(\Gamma_2/2 +\Gamma_{12}/2)^2}
\nonumber \\
& \simeq & 
{e\over h}\int d\omega 
 {f_L-f_R\over (\omega-\epsilon^{(1)}_{l_1})^2+(\Gamma_1/2)^2}
{\Gamma_2\Gamma_1|t|^2\over (\omega-{\epsilon}^{(2)}_{l_2}-
{|t|^2 (\omega-{\epsilon}^{(1)}_{l_1})\over 
(\omega-{\epsilon}^{(1)}_{l_1})^2+(\Gamma_1/2)^2}
)^2+({\Gamma_2\over 2}+{\Gamma_1\over 2}
{|t|^2 \over (\omega-{\epsilon}^{(1)}_{l_1})^2+(\Gamma_1/2)^2})^2
},
\label{eq:curr3}
\end{eqnarray}
where we have used definitions $i{\bf \Gamma}_{12}=
(\delta{\bf \Sigma}_{2})^r-(\delta{\bf \Sigma}_{2})^a$ and $\delta \epsilon^{(2)} = \Re (\delta{\bf \Sigma}_{2})^r$. 
Near resonance,
$\mu_L >\epsilon^{(1)}_{l_1} \sim \epsilon^{(2)}_{l_2}
>\mu_R$,
$\Gamma_{12}= \Gamma_1
{|t|^2 \over (\omega-\epsilon^{(1)}_{l_1})^2+(\Gamma_1/2)^2}$;
$\delta \epsilon^{(2)} = {|t|^2 (\omega-{\epsilon}^{(1)}_{l_1})\over 
(\omega-{\epsilon}^{(1)}_{l_1})^2+(\Gamma_1/2)^2}$.

\section{Nonequilibrium Distributions in the Dots and Population Inversion}
\label{sec:inversion}

Before discussing the resonant current, let us first consider the 
 {\it nonequilibrium} distributions of
electrons in the two dots. The distribution at each level
$l_i$ of dot $i$ is given by 
\begin{eqnarray}
F^i_{l_i}(\epsilon) &=&  [G^{+-}_{ii}]_{l_il_i}(\epsilon)
/([G^{r}_{ii}]_{l_il_i}-[G^{a}_{ii}]_{l_il_i})
\nonumber\\ 
&\frac{HFA}{i=2,l_i=l_2}&\rightarrow  
{f^{FD}_R + f^{FD}_L\lambda(\epsilon^{(2)}_{l_2}) \Gamma_1/\Gamma_2 
\over 1+ \lambda(\epsilon^{(2)}_{l_2})\Gamma_1/\Gamma_2 }
\nonumber \\
&\frac{HFA}{i=1,l_i=l_1}&\rightarrow  
{f^{FD}_L + f^{FD}_R\lambda(\epsilon^{(1)}_{l_1}) \Gamma_2/\Gamma_1 
\over 1+ \lambda(\epsilon^{(1)}_{l_1})\Gamma_2/\Gamma_1 
}
\end{eqnarray}
with $\lambda(\epsilon^{(i)}_{l_i}) \Gamma_{\bar{\imath}}= 
\Gamma_{\bar{\imath}}|t|^2 \sum_{klm} 
\left[ {\bf \tilde{g}}^r_{\bar{\imath}}\right]_{kl}
\left[ {\bf \tilde{g}}^a_{\bar{\imath}}\right]_{lm}$ 
 the energy level
broadening due to tunneling between dots.
When the level $\epsilon^{(2)}_{l_2} < \mu_R$ ($\mu_R <\mu_L$
according to our convention), 
$f^{FD}_L(\epsilon^{(2)}_{l_2})=f^{FD}_R(\epsilon^{(2)}_{l_2})=1$, 
$F^2_{l_2}(\epsilon^{(2)}_{l_2})=1$, this level in the dot
is occupied; when 
the level $\epsilon^{(2)}_{l_2} > \mu_L$, 
$f^{FD}_L(\epsilon^{(2)}_{l_2})=f^{FD}_R(\epsilon^{(2)}_{l_2})=0$, 
$F^2_{l_2}(\epsilon^{(2)}_{l_2})=0$, this level is empty.
>From Eq.(\ref{eq:curr2}), 
the energy levels in these two cases do not contribute
to the resonant tunneling. A more interesting case
is for levels lying between $\mu_R$ and $\mu_L$,
where $f^{FD}_L=1$ and $f^{FD}_R=0$,
electrons are in nonequilibrium states, and $F^2_{l_2}\sim 
{\lambda_2\Gamma_1\over \Gamma_2+\lambda_2\Gamma_1}$
strongly depends on the energy spectrum in the first
dot. Using the properties of $\left[ {\bf \tilde{g}}^r_1\right]_{l_il_j}$,
when a level-$i_u$ of dot-2 is in resonance with a level
in dot-1, $\lambda_2 \sim |t|^2/(4\Gamma_1^2)$,
the occupation number of level-$i_u$ is 
$F^2_{i_u} \sim 
{|t|^2 \over 4\Gamma_1\Gamma_2+|t|^2}$;
when a level-$i_d$ of dot-2 is not in resonance with levels in dot-1, 
$\lambda_2 \sim (|t|/\Delta E)^2 \ll 1$,
the occupation number of level-$i_d$ is given by
$F^2_{i_d} \sim 
|t/\Delta E|^2 \Gamma_1/\Gamma_2 \ll \Gamma_1/\Gamma_2$.

There is a {\it population inversion}
in the dot-2 ($|t| \sim \Gamma$)
when $\epsilon^{(2)}_{i_u} > \epsilon^{(2)}_{i_d}$:
the population inversion in dot-2 can be achieved due
to the resonance of a higher level-$i_u$ 
(relative to an active level-$i_d$) in dot-2 with a
level in dot-1.
The nonequilibrium electron distribution in dot-1 is similar.
When the level $\epsilon^{(1)}_{l_1} < \mu_R$, 
$F^1_{l_1}(\epsilon^{(1)}_{l_2})=1$, this level in the dot
is occupied; when 
the level $\epsilon^{(1)}_{l_1} > \mu_L$, 
$F^1_{l_1}(\epsilon^{(1)}_{l_1})=0$, this level is empty.
When $\mu_R < \epsilon^{(1)}_{l_1} < \mu_L$,
$F^1_{l_1}\sim 
{\Gamma_1\over \Gamma_1+\lambda_1\Gamma_2}$. When this level
is not in resonance with a level in dot-2, 
$\lambda_1 \sim (|t|/\Delta E)^2 \ll 1$, $F^1_{l_1}\sim 1$;
otherwise, $\lambda_1 \sim |t|^2/(4\Gamma_2^2)$,
$F^1_{l_1}\sim 1/(1+|t|^2/(4\Gamma_1\Gamma_2))$. Thus, with
$|t|\ll \Gamma_i$, $F^1_{l_1} \sim 1$ when 
$\mu_R < \epsilon^{(1)}_{l_1} < \mu_L$.

This population inversion is similar to that in quantum
cascade lasers for superlattices \cite{capasso94} and
other proposals for double quantum wells \cite{goldman91}. 
However, the population inversions proposed previously for quantum wells
are achieved without elastic resonant tunneling. 
For example, in the double quantum well systems \cite{goldman91}, 
the upper level
is attached to the lead with high chemical potential $\mu_h$ and
with $E_u < \mu_h$, so the upper level is occupied;
the lower level (either in the same well or in another well)
is attached to the lead with lower chemical potential $\mu_l$ with
$E_l > \mu_l$, thus it is empty. 
Since transverse momentum is conserved and can be neglected
the quantized energy levels in the wells
are mismatched, there is {\it no} elastic resonant tunneling,
and the system is in quasi equilibrium with two chemical 
potentials $\mu_h$ and $\mu_l$. The electrons can only tunnel
via emitting photons. In the population inversion we proposed above,
the energy levels are not in quasi equilibrium with the
leads. The population inversion is achieved through resonant tunneling.
Comparing to the proposal in Ref.\cite{goldman91} for double quantum
wells, we believe that our proposal is easier to realize practically.

In the above analysis, we have neglected the inelastic scattering
effects due to e-e and e-p couplings. In real systems,
the e-e and e-p scatterings have important effects on the
nonequilibrium distribution of electrons in the dots.
At high temperatures, the nonequilibrium distribution
will be thermalized 
due to inelastic relaxation. 
Here we concentrate on low temperatures,
when the dominant contribution to $\bbox{\sigma}$ comes 
from higher order e-e correlations which will affect the
NGF's. Taking this fact into account, 
the $F$-function with a finite $\bbox{\sigma}$ becomes:
\begin{equation}
F^2_{i}(\epsilon^{(2)}_{l_2}) 
={\sigma_{22}^{+-}+ \lambda(\epsilon^{(2)}_{l_2}) \Gamma_1 
\over \Gamma_2+ \gamma_2+\lambda(\epsilon^{(2)}_{l_2})\Gamma_1  
} ,
\end{equation}
where $i\bbox{\gamma}_{j}=
\bbox{\sigma}_{j}^r-\bbox{\sigma}_{j}^a$ with the
assumption $[\bbox{\gamma}_j]_{l_jl'_j}=\gamma_2$. Physically, $\gamma_2$
is just the average level broadening due to e-e scattering. $\sigma_{22}^{+-}$
is responsible for changing of the distribution functions. 
Note that $\sigma_{22}^{+-} \alt \gamma_2$.
Here $\lambda(\epsilon^{(2)}_{l_2})$ 
has been changed due to the introduction of
$\sigma$. So at resonance of level-$i_u$, 
$F^2_{i_u} \sim 
{{|t|^2 \Gamma_1+4(\sigma_{22})_{i_u,i_u}^{+-}(\Gamma_1+\gamma_1)^2
\over  
4(\Gamma_2+\gamma_2) (\Gamma_1+\gamma_1)^2+|t|^2\Gamma_1} }$, while
$F^2_{i_d} \sim (\sigma_{22})_{i_d,i_d}^{+-}/(\Gamma_2+\gamma_2)$.
If $(\sigma_{22})_{i_u,i_u}^{+-} \sim (\sigma_{22})_{i_d,i_d}^{+-}$,
it is easy to show that $F^2_{i_u}>F^2_{i_d}$. In the extreme case
that $(\sigma_{22})_{i_d,i_d}^{+-} \sim \gamma_2$ and 
$(\sigma_{22})_{i_u,i_u}^{+-} \sim 0$,
the population inversion
survives when $|t|^2 >{4\gamma_2 (\Gamma_2+\gamma_2)
\over \Gamma_1\Gamma_2}(\Gamma_1+\gamma_1)^2$.
For the e-e scattering in the typical experiments
$\gamma_i \alt \Gamma_{i}$ \cite{lihong}, so the population
inversion can be achieved at $|t|\gg \Gamma$. 
Consequently, the e-e correlations reduce the magnitude
of the resonant current but do not qualitatively change the resonance peak
structure.

\section{Nonequilibrium Currents and Peak Splittings}
\label{sec:exp}

\subsection{Nonequilibrium Currents}
In this section, we discuss the recent nonequilibrium experiments of Ref.\cite{vaart95} using the general equations derived above.
In these experiments, $\Gamma_i$, $|t| \ll \Delta E$, so we can neglect the contributions
of energy levels far from resonance (i.e. for experiment \cite{vaart95},
we only need take into account 4 relevant levels). We use Eq.(\ref{eq:curr2})
to calculate the currents, which neglects the high order scattering-in(out)
processes. This is a reasonable approximation for low temperature
experiments of quantum dots with electron numbers on
the order of 90 \cite{lihong}, as we discussed above.

\begin{figure}
\centerline{
 \epsfxsize=14cm \epsfbox{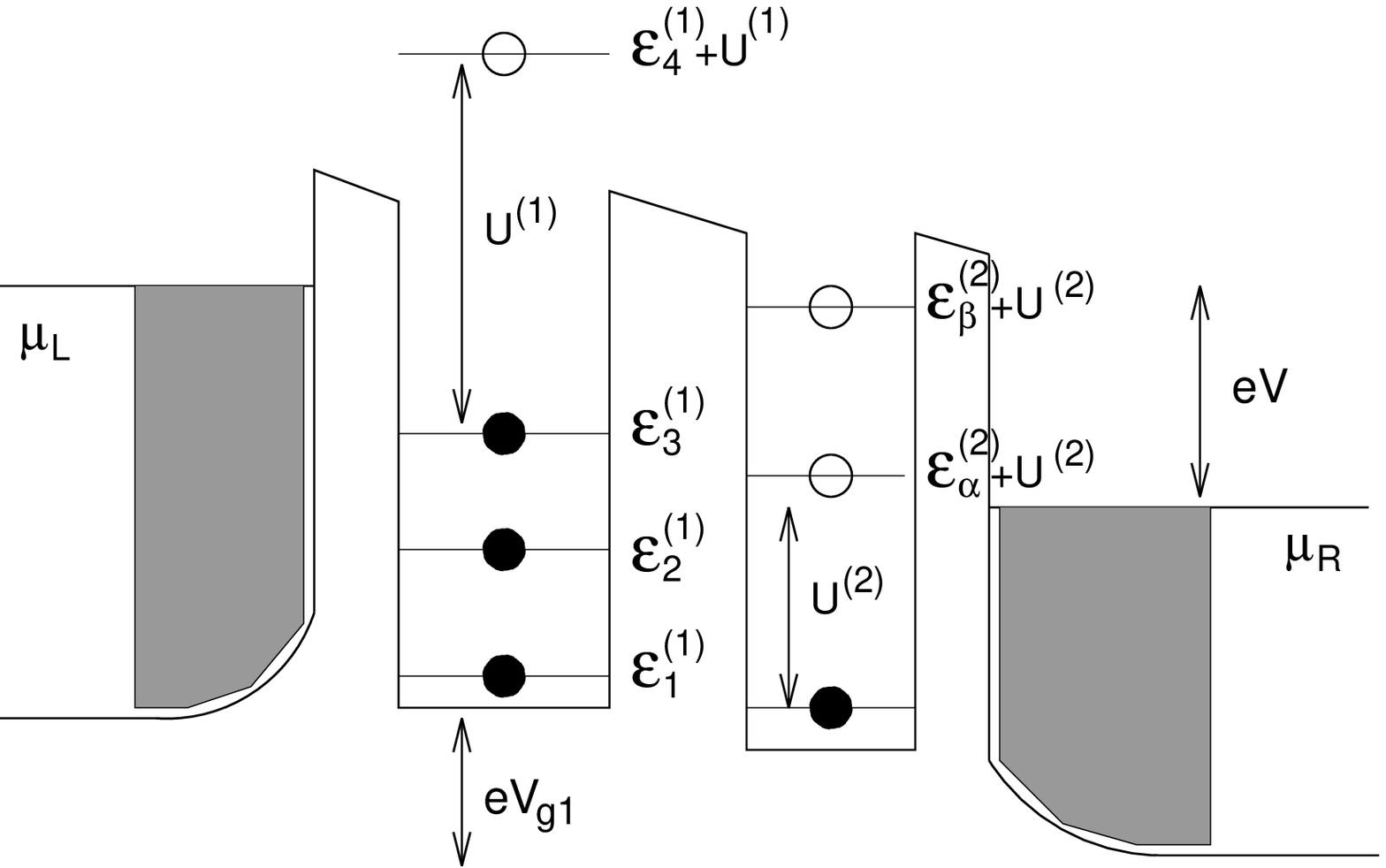}
}
\vskip 0.0cm
\caption{ Schematic energy potential landscape of the double
quantum dots and leads. $\epsilon^{(i)}_{l_i}$'s
($i=1, 2$) are the energy spectra in dot-1 and dot-2, respectively.
The solid and empty dots represent populated and unpopulated 
states. 
$U^{(i)}$, a charging energy, produces populated states
separated from the empty states. $V_{g1}$ is a gate voltage
applied to dot-1. $V$ is a bias voltage.  
 \label{fig:engy}}
\vskip 1in
Fig.1 of PRB, Zang et al
\end{figure}

\begin{figure}
\centerline{
 \epsfxsize=14cm \epsfbox{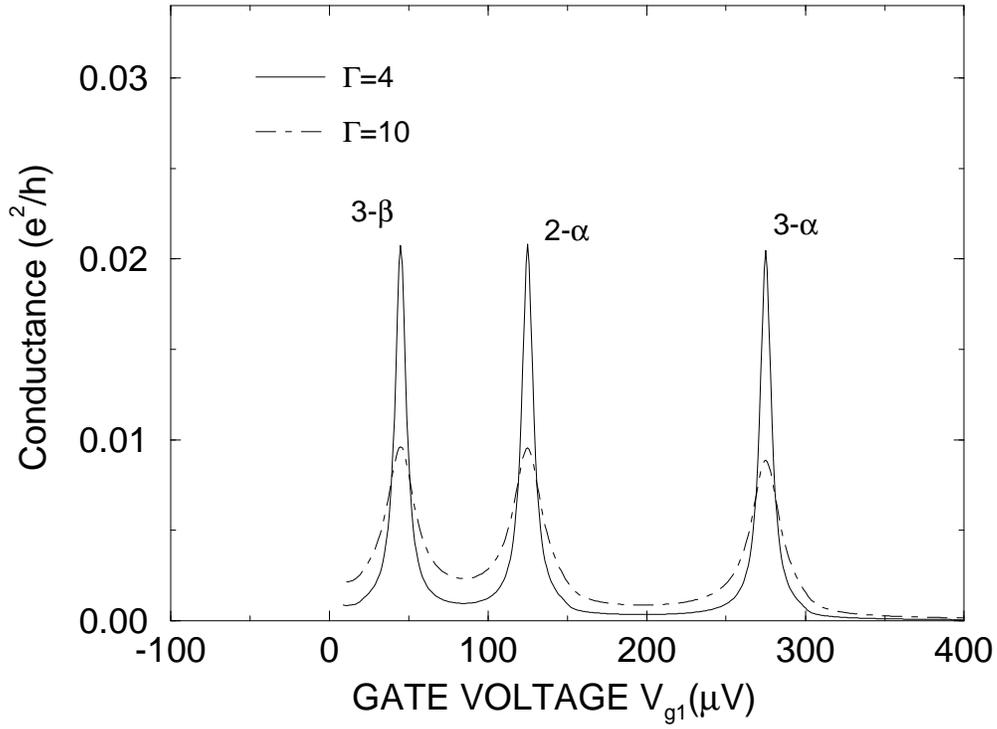}
}
\caption{ 
Current through the DQD vs. gate voltage
$V_{g1}$ using a bias voltage $V=280\mu V$. 
$|t|=0.2\mu eV$; $\Delta E^{(1)}=150\mu eV$;
$\Delta E^{(2)}=230\mu eV$
\label{fig:iv}} 
\vskip 1in
Fig.2 of PRB, Zang et al
\end{figure}

The oscillations of the nonlinear resonant tunneling current 
vs. gate voltage in experiment \cite{vaart95} can be
readily explained if we take into account the charging 
effect \cite{lee-dot,rev} for 
the renormalized HF energy levels $\epsilon^{(1)}_{l_1}$ and $\epsilon^{(2)}_{l_2}$ 
[see Fig.\ref{fig:engy}]. In Fig.\ref{fig:iv} we plot the
conductance of the current tunneling through the DQD vs. the
gate voltage $V_{g1}$ on dot-1 using a constant bias voltage of $280 \mu V$.
(Only  one group of the peaks is shown). 
The current is calculated
using Eq.(\ref{eq:curr2}) 
for $|t|=0.2\mu eV$; 
$\Gamma_1=\Gamma_2=4.0\mu eV$, 10.0$\mu eV$;
 the level spacing in dot-1 and 
dot-2 as $\Delta E^{(1)} = 150\mu eV$ and 
$\Delta E^{(2)} = 230\mu eV$, respectively. 

Our assignment of the peaks in Fig.(\ref{fig:iv}) is the same
as that in Ref.\cite{vaart95}. 
We assume in Fig.(\ref{fig:iv}) that the spectrum in dot-2 is 
fixed while varying $V_{g1}$. Using the energy level spacings given in 
Ref.\cite{vaart95}, the spacing between the first and second peaks
is $\sim 80 \mu eV$, and the spacing between the second and
third peaks is $\sim 150 \mu eV$. However, in the experimental
I-V curves, the first spacing 
is larger than the second one. 
One possible reason is that the level spacing is not regular,
which is suggested in Ref.\cite{vaart95}.
Note that the gate voltage scale
for these spacings are $\sim 900\mu V$ and $2500\mu V$ respectively.
In Ref.\cite{vaart95} they are converted to energy scales
as $70\mu eV$ and $200\mu eV$. 
Then the level spacing between level-2 and level-3 in dot-1
is barely half of the average level spacing. 
Another possibility is that this discrepancy can be due
to the charging effect. 
As discussed in Sec.\ref{sec:inversion}, without resonance between
the two dots, the 
 ``active'' levels in dot-1 are filled and the ``active'' levels
in dot-2 are empty; with resonance, the resonance level in
dot-2 has population $\sim |t|^2/(4\Gamma_1\Gamma_2)$ while
the resonance level in dot-1 has population 
$\sim 1-|t|^2/(4\Gamma_1\Gamma_2)$ ($|t|\ll\Gamma_i$). 
So the levels in dot-1 are shifted {\it down} by $\delta \epsilon_1
\propto |t|^2/(\Gamma_1\Gamma_2)$ and the levels in
dot-2 are shifted {\it up} by $\delta \epsilon_2
\propto |t|^2/(\Gamma_1\Gamma_2)$. The net effect is that
the current peak positions are pushed {\it upward} by $\delta V \propto |t|^2/(\Gamma_1\Gamma_2)$. Recall that the peak height for
the resonant current is $I \propto |t|^2/(\Gamma_1\Gamma_2)$. 
In the experimental I-V curve, the magnitude of the
first peak is much smaller than that of the
second and third peaks, so the second and third peaks are
shifted {\it up} much more than the first peak (due to large charging
effects). Thus the first peak spacing can be larger than the second
one even if we assume the active energy levels are close
to their average values.

Another consequences of this charging
effect is that the Lorentzian line shape of the current will
be changed. For the experiments in \cite{vaart95}, the peak should
be distorted slightly to the right side due to charging effects alone
(e.g. the high gate voltage side has a longer tail).
However, couplings to the other levels will induce asymmetric shoulders
as observed in the experiments.
Note that there is no 4-$\alpha$
peak [see Fig.\ref{fig:engy}]. The absence of this 4-$\alpha$
peak was argued in \cite{vaart95} to be due to rapid relaxation
of level-4 to level-3. As we discussed in Sec.\ref{sec:inversion}
 that {\it if} the energy of level-4
is below $\mu_L$, then level-3 is already filled without
coupling between level-4 and level-3, so even at high temperature,
there should be no substantial relaxation from level-4 to level-3.
Since the experiments are conducted at very low temperature (34mK),
 the argument that
{\it when level-4 is lower than $\mu_L$}
there is large relaxation from level-4 to
level-3  is questionable.  From our Fig.\ref{fig:engy}, we can see
that, since all the levels below $\mu_L$ in dot-1 are filled, 
they is a large charging energy between 4-$\alpha$ peak and this group
of peaks, so the 4-$\alpha$ peak is ``missing''. The authors in 
\cite{vaart95} found that the amplitude of the
resonance peak will stay constant with increasing bias voltage,
which was argued to support the
 large relaxation of level-4 to level-3.
However, as we
discussed in Sec.\ref{sec:inversion}, the 
population of the lowest active level stays constant 
with increasing 
bias voltage $V$ even without inelastic relaxations, 
the amplitude of the
resonance will stay constant with increasing of bias voltage,
so this phenomenon cannot tell if there is large relaxation.
Thus, all the observations in Ref.\cite{vaart95}
can be well explained by our calculations without 
the assumption of large inelastic relaxation at low temperature
$\sim 34mK$.

From our calculations the half widths
of these peaks are not $\sim\Gamma/2$, due to
coupling between multi-levels in each dot,
 and the increase of
$\Gamma$ will eventually destroy the resonance as shown in Fig.(\ref{fig:iv}). 

\subsection{Peak Splittings}

\begin{figure}
\centerline{
 \epsfxsize=14cm \epsfbox{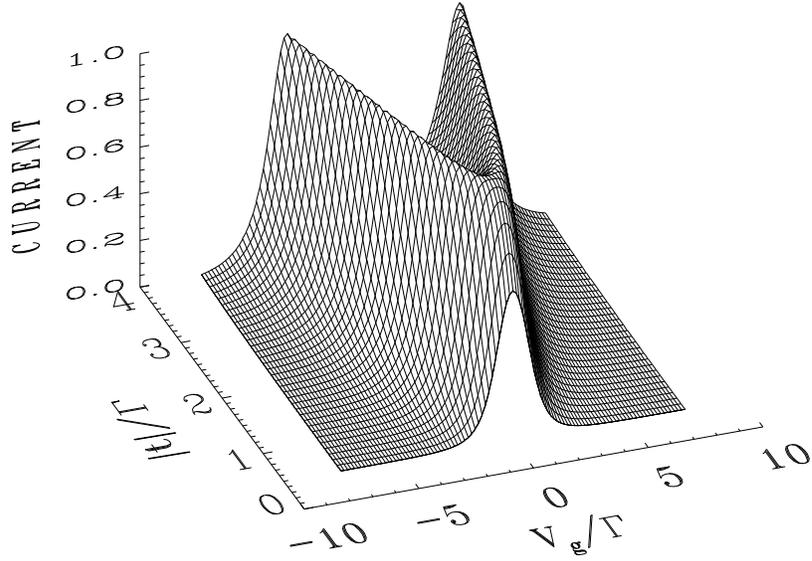}
}
\caption{  
Tunneling current peak splitting vs. coupling 
$|t|$ and varying gate voltage $V_g$ for a 
symmetrical double-dot: $\Gamma_1=\Gamma_2\equiv\Gamma$
and $\epsilon^{(1)}_1=\epsilon^{(2)}_2$.
\label{fig:splt}}
\vskip 1in
Fig.3 of PRB, Zang et al
 \end{figure}

In a symmetrical geometry, $\Gamma_1=\Gamma_2=\Gamma$
and $\epsilon^{(1)}_{l_1}=\epsilon^{(2)}_{l_2}=\epsilon$, 
the resonant tunneling
current Eq.(\ref{eq:curr3}) can be simplified as following:
\begin{eqnarray}
J \simeq {e\over h}\int d\omega 
 (f_L-f_R)
{\Gamma^2|t|^2\over [(\omega-\epsilon)^2-(|t|^2-\Gamma^2/4)]^2+
\Gamma^2|t|^2}.
\label{eq:curr-s1}
\end{eqnarray}
It is readily seen that there will be a splitting
in the resonance peak $2(|t|^2-\Gamma^2/4)^{1/2}$
when $|t| > \Gamma/2$. The $|t|$-dependence
of the resonant peak splitting is shown in Fig.(\ref{fig:splt}).
The physical meaning of this splitting is clear in the
strong interdot coupling limit $|t|\gg\Gamma$: 
\begin{eqnarray}
J \simeq {e\over h}\int d\omega 
 (f_L-f_R)
\left[{\Gamma^2/4\over (\omega-\epsilon-|t|)^2+
\Gamma^2/4}+{\Gamma^2/4\over (\omega-\epsilon+|t|)^2+
\Gamma^2/4} \right],
\label{eq:curr-s2}
\end{eqnarray}
which is just resonant tunneling through a two level
single dot system. This is not surprising, since in
the strong interdot coupling limit, the two dots can be treated
as a single system (for these two resonant levels--this
is not true for other non-resonant levels),
two degenerate levels will be split by $2|t|$
due to interdot coupling.
However, if the dot-lead coupling strength is stronger,
then each dot is coupled to its own reservoir, and then
weakly coupled to each other. If the two reservoirs are
out of equilibrium, then these two dots cannot be treated
as a single system.
From Eq.(\ref{eq:curr-s1}), there will be no peak splitting in this regime \cite{note2}. 
Notice that, for $|t| \agt \Gamma/2$, the peak splitting predicted here
is different from that expected from a naive argument based on
degeneracy lifting \cite{waugh95}.
When $|t| \gg \Gamma$, the splitting $\sim 2|t|$, and
our result becomes equivalent to that in CBT.

\section{Summary}
\label{sec:sum}

In summary, we have developed a microscopic theory
for the equilibrium and nonequilibrium transport through  
a double quantum dot (DQD). The analytical expressions
for resonant tunneling current peaks were derived in both 
linear and nonlinear conductance regimes. We showed that the
nonequilibrium electron distribution has important effects
on the resonant tunneling through DQD systems in the nonlinear
regime, and it is necessary to be take into account 
these nonequilibrium distributions to interpret nonlinear 
tunneling experiments. We also find that multi-level coupling 
affects the width of resonant tunneling current peaks.
Using the exact results for a non-interacting system,
we obtained a tunneling peak splitting for a symmetrical double dot
$2\sqrt{|t|^2-(\Gamma/2)^2}$ when $|t|\geq \Gamma/2$. 
Most interestingly, we predict that the population of 
electrons in the active levels of one dot can be inverted 
in the nonequilibrium regime by changing the gate voltage 
to make the upper level in the 2nd dot match
with an active level in the 1st dot. This mechanism 
of population inversion could be used in a double quantum
well resonant tunneling diode. 
The population inversion proposed here is different from that
of QCLs \cite{capasso94}, which is
based on photon-assisted tunneling in a superlattice
\cite{kazarinov} or a double quantum well \cite{goldman91}.

{\bf Acknowledgements:} We thank S. Trugman 
for helpful discussions. Work at Los Alamos
is performed under the auspices of the U.S. DOE.

\newpage

\newpage

\begin{thebibliography}{10}
\bibitem{note3} In cases $\Re \bbox{\sigma}$ is not diagonal in the
$\{\Psi_i\}$ space, one need to re-define the basis states, then current
equation is still given by Eq.(\ref{eq:curr2}).
\bibitem{note0} When $|t|\agt\Gamma$, the population in the
resonance level of dot-1 will be reduced near resonance, while that in dot-2
will increase.
\bibitem{note2} In Fig.2 of Ref.\cite{waugh95}, 
the authors also observed peak splitting
for small inter-dot conductance. It is difficult to judge whether 
$|t|<\Gamma/2$ or not. From Fig.3 of Ref.\cite{waugh95}, the smallest
$F\sim 0.05$, which gives peak splitting $\sim 10\mu eV$ ($<|t|$ according to
our calculation).
Considering $\Delta E\sim 30\mu eV$, $\Gamma$ should be smaller
than $10\mu eV$ to observe resonance. A reasonable interpretation is that
for the smallest peak splitting in \cite{waugh95} the inter-dot
tunneling is  $|t| \agt \Gamma$.
\end{thebibliography}
\end{document}